\documentclass[final,5p,times,twocolumn]{elsarticle}

\usepackage{color,soul}
\usepackage{graphicx}
\usepackage{amssymb}
\usepackage{amsmath}
\biboptions{square,comma}
\usepackage{caption}
\journal{Journal of Nuclear Materials}

\newcommand{\xolotlpsi}{{\small XOLOTL-PSI}}
\newcommand{\ovito}{{\small OVITO}}
\newcommand{\lammps}{{\small LAMMPS}}
\newcommand{\ksome}{{\small kSOME}}
\newcommand{\scidavis}{{\small SCIDAVIS}}
\newcommand{\azero}{$a_0$}
\newcommand{\EMD}{$E_{MD}$}
\newcommand{\Ed}{$E_{d}$}

\newcommand{\Estar}{$E^{*}$}

\newcommand{\Tm}{$T_m$}

\newcommand{\NF}{$N_{F}$}
\newcommand{\one}{$^1$}

\newcommand{\etal}{{\it et al}}

\newcommand{\Smaxsia}{$S_{SIA}^{max}$}
\newcommand{\Smaxvac}{$S_{vac}^{max}$}
\newcommand{\Nlargesia}{$C_{SIA}^{large}$}
\newcommand{\Nlargevac}{$C_{vac}^{large}$}
\newcommand{\gsr}{$g_s(r)$}

\newcommand{\onehalf}{$1/2$}
\newcommand{\twothird}{$2/3$}

\begin{document}
\begin{frontmatter}

\title{Displacement cascades and defects annealing in tungsten, Part I: defect database from molecular dynamics simulations}

\author{Wahyu Setyawan$^{1,*}$, Giridhar Nandipati\one, Kenneth J. Roche$^{1,2}$, Howard L. Heinisch\one, Brian D. Wirth$^{3,4}$ and Richard J. Kurtz\one}
\cortext[]{Corresponding author. Tel.: +1(509)371-7692, fax +1(509)375-3033, {\it E-mail address}: wahyu.setyawan@pnnl.gov}
\address{
\one\ Pacific Northwest National Laboratory, Richland, WA 99352, USA\\
$^2$ Department of Physics, University of Washington, Seattle, WA 98195, USA\\
$^3$ Department of Nuclear Engineering, University of Tennessee, Knoxville, TN 37996, USA\\
$^4$ Oak Ridge National Laboratory, Oak Ridge, TN 37831, USA
}

\begin{abstract}
Molecular dynamics simulations have been used to generate a comprehensive database of surviving defects due to displacement cascades in bulk tungsten. Twenty one data points of primary knock-on atom (PKA) energies ranging from 100 eV (sub-threshold energy) to 100 keV ($\sim$780$\times$\Ed, where \Ed\ is the average displacement threshold energy) have been completed at 300 K, 1025 K and 2050 K. Within this range of PKA energies, two regimes of power-law energy-dependence of the defect production are observed. A distinct power-law exponent characterizes the number of Frenkel pairs produced within each regime. The two regimes intersect at a transition energy which occurs at approximately 250$\times$\Ed. The transition energy also marks the onset of the formation of large self-interstitial atom (SIAs) clusters (size 14 or more). The observed defect clustering behavior is asymmetric, with SIA clustering increasing with temperature, while the vacancy clustering decreases. This asymmetry increases with temperature such that at 2050 K ($\sim$0.5\Tm) practically no large vacancy clusters are formed, meanwhile large SIA clusters appear in all simulations. The implication of such asymmetry on the long-term defect survival and damage accumulation is discussed. In addition, rare $<$100$>$\{110\} SIA loops are observed.
\end{abstract}

\end{frontmatter}

\section{Introduction}
One of the challenging issues in fusion materials is determining the effects of neutron damage in tungsten plasma facing components under fusion relevant conditions. The challenge stems from the unavailability of a 14-MeV fusion neutron source. Therefore, development of models to predict neutron damage through computer simulations is essential for designing and interpreting experiments performed in fission reactors and linear plasma devices. An example is the extensive effort to develop an open-source code, \xolotlpsi\ \cite{xolotl}, to simulate plasma-surface interactions with tungsten. Being a multi-scale simulation tool, \xolotlpsi\ requires inputs from lower-level atomistic simulations. In this effort, a recently formulated W potential \cite{JuslinHeWpot} is adopted to gather knowledge of atomistic processes for \xolotlpsi.  This paper is concerned with the development of a database of primary defect states of radiation damage in tungsten. Along the way, several discoveries are made including a transition energy that separates different regimes of energy associated with different defect survival mechanisms, as well as asymmetric defect clustering behavior (i.e. strong differences in self-interstitial atom vs. vacancy clustering).

Cascade damage databases at 300, 1025 and 2050 K are compiled. A systematic study of cascade energy ranging from 100 eV to 100 keV is performed at each temperature to better sample the full recoil energy spectrum of the primary-knock-on atom (PKA) due to fusion neutrons. As such, this paper presents the most comprehensive molecular dynamics (MD) database of cascade simulations in tungsten to date \cite{JuslinPhilMag10, FikkarJNM09, BjorkasNIMB09, ParkNIMB07}. The range of PKA energies covers sub-threshold, low- and high-energy regimes. Data at 2050 K allows the investigation of the influence of temperature on defect production and morphology to be performed up to 0.5 of the melting temperature, \Tm. With this comprehensive set of PKA energies and temperatures, we are able to reveal the profound effect of temperature on the asymmetry of defect clustering.

We have performed both displacement cascade and defect annealing simulations in tungsten. The results are presented in two parts. This paper reports on the primary damage production. Molecular dynamics simulations, with the \lammps\ code \cite{LAMMPS}, have been used to perform the cascade simulations. A companion paper focuses on modeling the long-term evolutions of the resulting defect structures using an object kinetic Monte Carlo (OKMC) technique. That research involved both the development of a modular and versatile OKMC software, \ksome\ and the simulations to determine the fate of the cascade damage, both of which are presented in a companion paper as Part II \cite{companion}.

\section{Methods}
The W potential used in this study \cite{JuslinHeWpot} is derived from Ackland-Thetford potential \cite{AcklandPM87}. A new modification hardens the potential at short distances for radiation damage simulations and improves the potential at distances of relevance to self-interstitial configurations. Prior to cascade simulations, all systems are thermalized for 30 ps at zero pressure with a Nos\'{e}-Hoover thermostat to obtain a proper distribution of atom positions and velocities. A cascade is initiated by giving a random PKA near the center of the simulation cell an initial velocity with a random direction. The PKA initial kinetic energy is denoted as \EMD. The cascade is simulated using a microcanonical ensemble for the first $\sim$10 ps and subsequently thermostated in a constant volume so that the target temperature is achieved within the next 1 ps. A typical total simulation time is $\sim$50 ps. The effect of the thermostat setting on defect counting and clustering is negligible. An adaptive time step is used, allowing a maximum displacement of 0.005 \AA\ per step. Sufficiently large simulation cells are used to ensure no displaced atoms cross the periodic boundaries. Displaced atoms are defined as those beyond 0.3\azero\ of any lattice site, where \azero\ is the lattice constant at the corresponding temperature. The calculated lattice constants at 300, 1025 and 2050 K are 3.17, 3.18 and 3.22 \AA, respectively. Note that these temperatures represent homologous temperatures of 0.07, 0.25 and 0.5 for the potential, respectively.

\begin{table}[htbp]
\caption{List of PKA energies (\EMD), simulation cells (cubes with side length $L$ expressed in lattice constant \azero) and the number of simulations ($N_r$).}
\begin{center}
\begin{tabular}{cccccc}
\hline \hline
\EMD & \EMD/\Ed & $L$ & $N_r$ &$N_r$&$N_r$\\
(keV) &  & $a_0$ & 300 K & 1025 K & 2050 K\\
\hline
0.1 & 0.78 & 15 & 40 & 40 & 40\\
0.15 & 1.17 & 15 & 40 & 40 & 40\\
0.2 & 1.56 & 15 & 40 & 40 & 40\\
0.3 & 2.34 & 15 & 40 & 40 & 40\\
0.5 & 3.91 & 20 & 20 & 20 & 20\\
0.75 & 5.86 & 20 & 20 & 20 & 20\\
1 & 7.81 & 30 & 20 & 20 & 20\\
1.5 & 11.72 & 30 & 20 & 20 & 20\\
2 & 15.63 & 30 & 20 & 20 & 20\\
3 & 23.44 & 30 & 20 & 20 & 20\\
5 & 39.06 & 40 & 20 & 20 & 20\\
7.5 & 58.59 & 40 & 20 & 20 & 20\\
10 & 78.13 & 50 & 20 & 15 & 15\\
15 & 117.19 & 50 & 20 & 15 & 15\\
20 & 156.25 & 64 & 20 & 15 & 15\\
30 & 234.38 & 64 & 20 & 15 & 15\\
40 & 312.50 & 64 & 15 & 15 & 15\\
50 & 390.63 & 80 & 15 & 15 & 15\\
60 & 468.75 & 80 & 15 & 15 & 15\\
75 & 585.94 & 100 & 15 & 15 & 15\\
100 & 781.25 & 120 & 20 & 20 & 20\\
\hline
 & 300 K & 1025 K & 2050 K & & \\
 \cline{2-4}
$a_0$ (\AA) & 3.167 & 3.184 & 3.216 & &\\
\hline \hline
\end{tabular}
\end{center}
\label{tableL}
\end{table}

As usual, a self-interstitial-atom (SIA) or a vacancy is determined from the Wigner-Seitz cellÕs occupancy. The list of PKA energies, simulation cell sizes and the number of simulation runs is presented in Table \ref{tableL}. When appropriate, the analysis presented in this paper is given as a function of PKA energy normalized by the average displacement threshold energy, \Ed. The normalization allows for better comparison with the results obtained for other materials and interatomic potentials. The conversion to the reduced energies is included in Table \ref{tableL} for convenience.

\section{Results}
\subsection{Defect production}
\begin{figure}[htbp]
\centering
\includegraphics[width=0.49\textwidth]{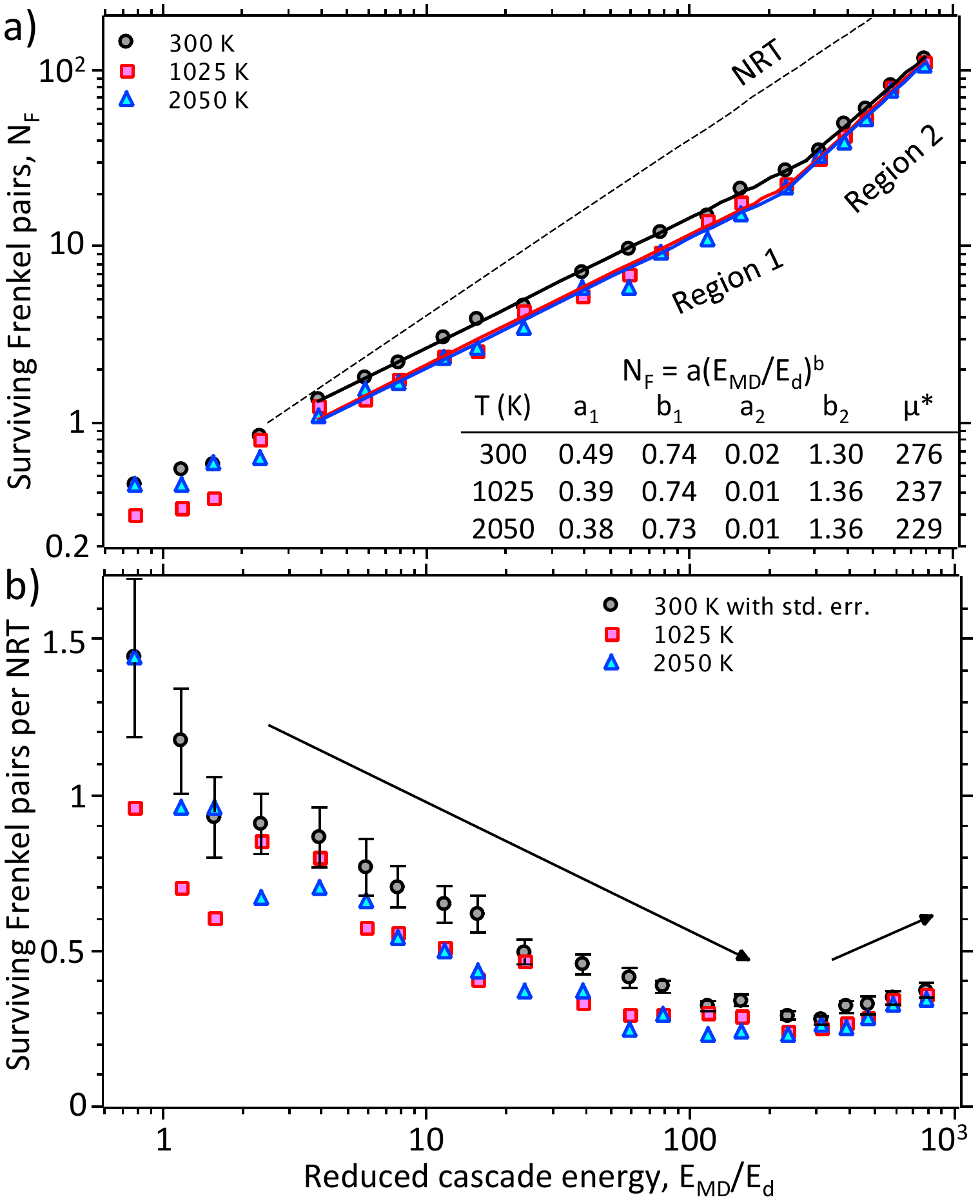}
\caption{a) Plots of \NF\ vs. \EMD/\Ed\ and power-law fits revealing two energy regimes with the transition occurring at a reduced energy $\mu^*$, b) Ratio of surviving MD Frenkel pairs to the number of displacements obtained from the NRT model.}
\label{figNF}
\end{figure}

The number of surviving Frenkel pairs, \NF, is plotted in Figure \ref{figNF}a as a function of reduced energy defined as \Estar\ $\equiv$ \EMD/\Ed, where the calculated average displacement threshold energy is \Ed\ = 128 eV \cite{JuslinHeWpot}. The value of \Ed\ is usually determined near zero K (in our case it is determined at 10 K) and signifies the energy at which the probability to create a Frenkel pair is 0.5. Comparing the value of \NF\ at \Estar\ = 1 for different temperatures, we observe the following. At 300 K, the value of \NF\ is $\sim$0.5, indicating that \Ed\ remains unchanged for a relatively wide range of low temperatures from 10 K to 300 K in tungsten. It is known that a thermally activated Frenkel pair may recombine hence increasing the effective value of \Ed. This phenomenon is likely responsible for the decrease of \NF\ to 0.3 at 1025 K. On the other hand, if the temperature is further increased, the increasing amplitude of atomic vibrations effectively weakens the atomic bonds and may lead to an increased probability of defect creation. The later phenomenon explains the increase of \NF\ back to 0.5 at 2050 K. Note that more simulations (40 runs) are performed in this sub-threshold regime to increase the statistical robustness. The standard error of \NF\ in this regime is 0.08, which indicates that the above variation of \NF\ is statistically meaningful. The error bars are omitted from Figure \ref{figNF}a for clarity, however they are included in the Supplemental Document \cite{supplemental}.

A steady creation of defects occurs for reduced damage energy \Estar\ $\ge$ 4. For this reason, only data points in this range are included in the power-law fits. The data naturally separates into two regimes, each of which is well fit as shown in Figure \ref{figNF}a. The lower- and higher-energy regimes are labeled Region 1 and Region 2, respectively. A linear regression is employed for the fitting. The highest energy data in Region 1 and the lowest energy data in Region 2 for the fitting are determined by minimizing the total residual of the two fit lines. The intersection of the fit lines is defined as the transition energy, $\mu$. The coefficients of the power-law curves and the reduced transition energy, $\mu^* \equiv \mu$/\Ed, are included in Figure \ref{figNF}a. Setyawan \etal. showed that in body-centered cubic (bcc) Cr, Fe, Mo and W near room temperature the transition occurs at $\mu^* \sim$ 280 to 380 \cite{CascadeTransition}.

In Figure \ref{figNF}a, the Norgett-Robinson-Torrens (NRT) displacement model, \NF\ = 0.4\Estar, is plotted for comparison. It is evident that the \NF\ curves obtained from MD simulations deviate from the NRT model. In Region 1, sublinear defect creation is observed and the slopes are practically invariant with respect to temperature ($b$ = 0.74). On the other hand, a superlinear behavior occurs in Region 2, for which the slope increases slightly from 1.30 at 300 K to 1.36 at both 1025 K and 2050 K. The defect production occurring during the MD simulations is significantly smaller than that predicted with the NRT model. Figure \ref{figNF}b shows the defect survival efficiency as a function of energy for the three temperatures. The plots exhibit a V-shape curve, reflecting this sublinear and superlinear behavior. Arrows in Figure \ref{figNF}b show this trend, with the minimum ratio of defect production from the MD simulations relative to the NRT model is found to be $\sim$0.25.

\subsection{Defect clustering}
Generally, clustering of defects is calculated with a relatively strict connectivity cutoff limited to the first (NN1) or the second (NN2) nearest neighbor distance. In MD cascade simulations, the distribution of defects in space is determined by the cascade process and migration of the defects within the simulation time. Self-interstitial atoms are mobile and tend to cluster due to mutual attraction between them. Therefore, it is appropriate to consider the cutoff based on the clustering behavior of an interstitial pair (di-interstitial). We find that the most stable dumbbell is a $<$111$>$ dumbbell and the most stable configuration for a di-interstitial is two parallel $<$111$>$ dumbbells separated at the third nearest neighbor distance (NN3). These results are consistent with a density functional theory (DFT) calculation \cite{BecquartJNM10}. 

\begin{figure}[htbp]
\centering
\includegraphics[width=0.48\textwidth]{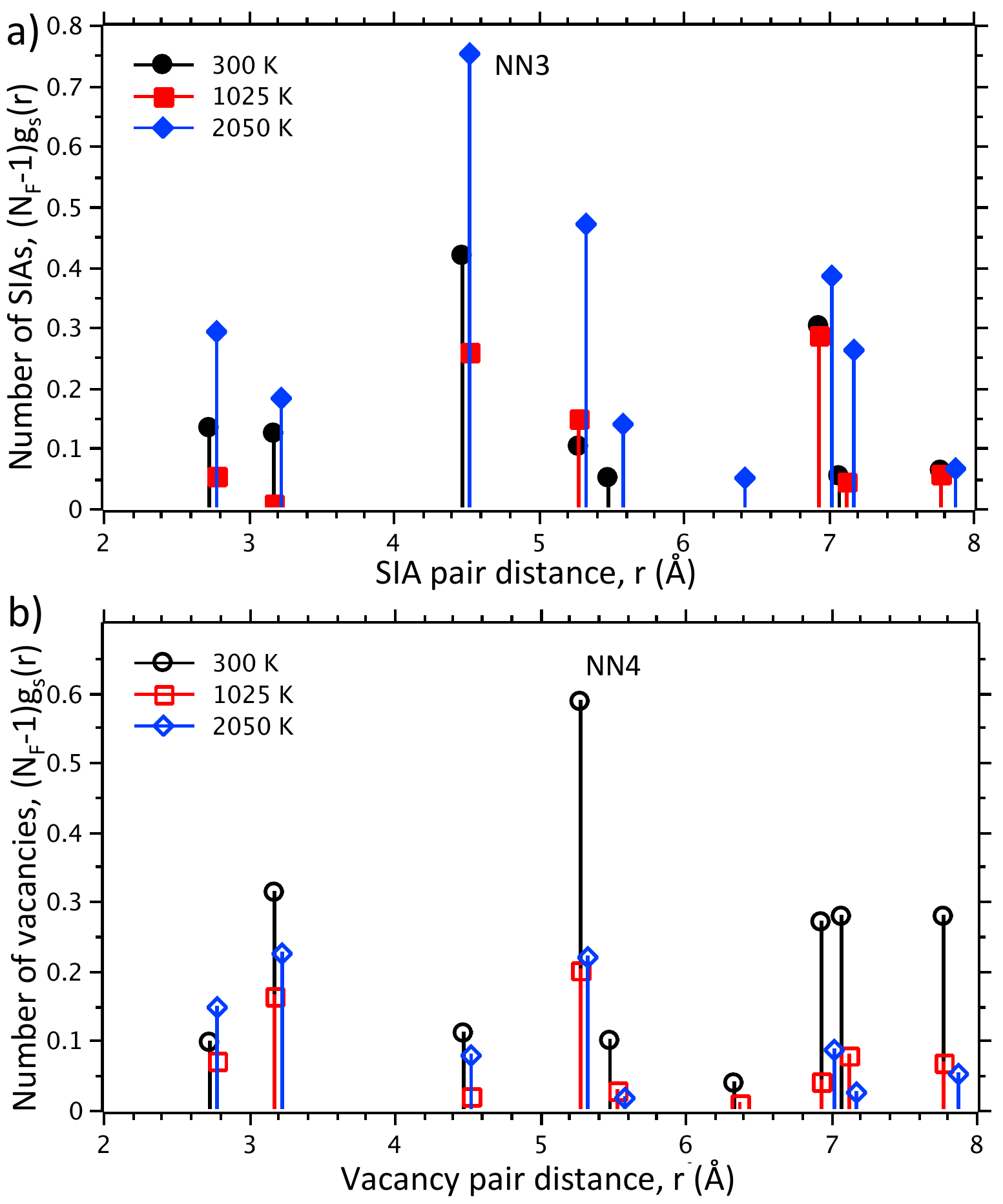}
\caption{{\it Sparse} radial distribution function, (\NF$-$1)\gsr\ (see Equation 1), for a) SIAs and b) vacancies. These distributions are taken from the 10-keV cascades and reveal a dominant correlation at the third (NN3) and the fourth (NN4) nearest neighbor distance for di-interstitials and di-vacancies, respectively.}
\label{figcdf}
\end{figure}

Unlike a di-interstitial, a DFT calculation shows that a di-vacancy is unstable \cite{BecquartNIMB07} (the strongest repulsion being at the NN2 separation). Note that a three-vacancy cluster is stable and can attract additional vacancies. Nevertheless, a vacancy is practically immobile at MD time scales. Therefore, it is desirable to use a cutoff that corresponds to potential for clustering if the cascades were to be annealed for a longer time in OKMC simulations. Such a cutoff can be deduced from a {\it sparse} radial distribution function, \gsr, defined as
\begin{align}
g_s(r) & \equiv \frac{2!(N-2)!}{N!} \sum\limits_{i=1}^{N-1} \sum\limits_{j>i}^{N} {\delta(r_{ij}-r)} \\
I_s(r) & = \int_0^r  g_s(x) dx, \text{\ \ \ } I_s(\infty) = 1
\end{align}
where $N$ is the number of defects of interest. The notation \gsr\ lends a similarity to the radial distribution function $g(r)$ but for a sparse distribution (hence the subscript $s$). With the above definition, a quantity ($N$$-$1)\gsr\ denotes the average number of neighboring point defects at a distance $r$ from a defect. The plots of (\NF$-$1)\gsr\  for SIAs and vacancies from the 10-keV cascades are shown in Figure \ref{figcdf}. Note that the plots for all energies in Region 1 are similar to those depicted in Figure \ref{figcdf}. It will be shown later that SIA loops and vacancy loops can form at energies in Region 2. The formation of such loops enhances the peak at NN1 of the (\NF$-$1)\gsr\  plots. Therefore, the plots from Region 1 are more appropriate to study the clustering tendency of individual point defects.

In Figure \ref{figcdf}, a dominant peak at NN3 is found for the SIAs, further reflecting the stable configuration of a di-intersitial as described before. On the other hand, for vacancies, a dominant spatial correlation is found at the fourth nearest neighbor distance (NN4). The correlation of vacancies at NN4 was also observed for displacement cascades in Fe \cite{StollerJNM97}. From this analysis, we take NN3 and NN4 as the clustering cutoff for SIAs and vacancies, respectively and use this as a default through the remainder of this article. However, for completeness, we generate another set of clustering data using a general cutoff of NN2 for both SIAs and vacancies, and this full analysis of the cluster counting data is compiled in the Supplemental Document \cite{supplemental}. Both sets are also studied in the OKMC simulations to investigate their respective long-time evolution \cite{companion}.

\begin{figure*}[htbp]
\centering
\includegraphics[width=0.95\textwidth]{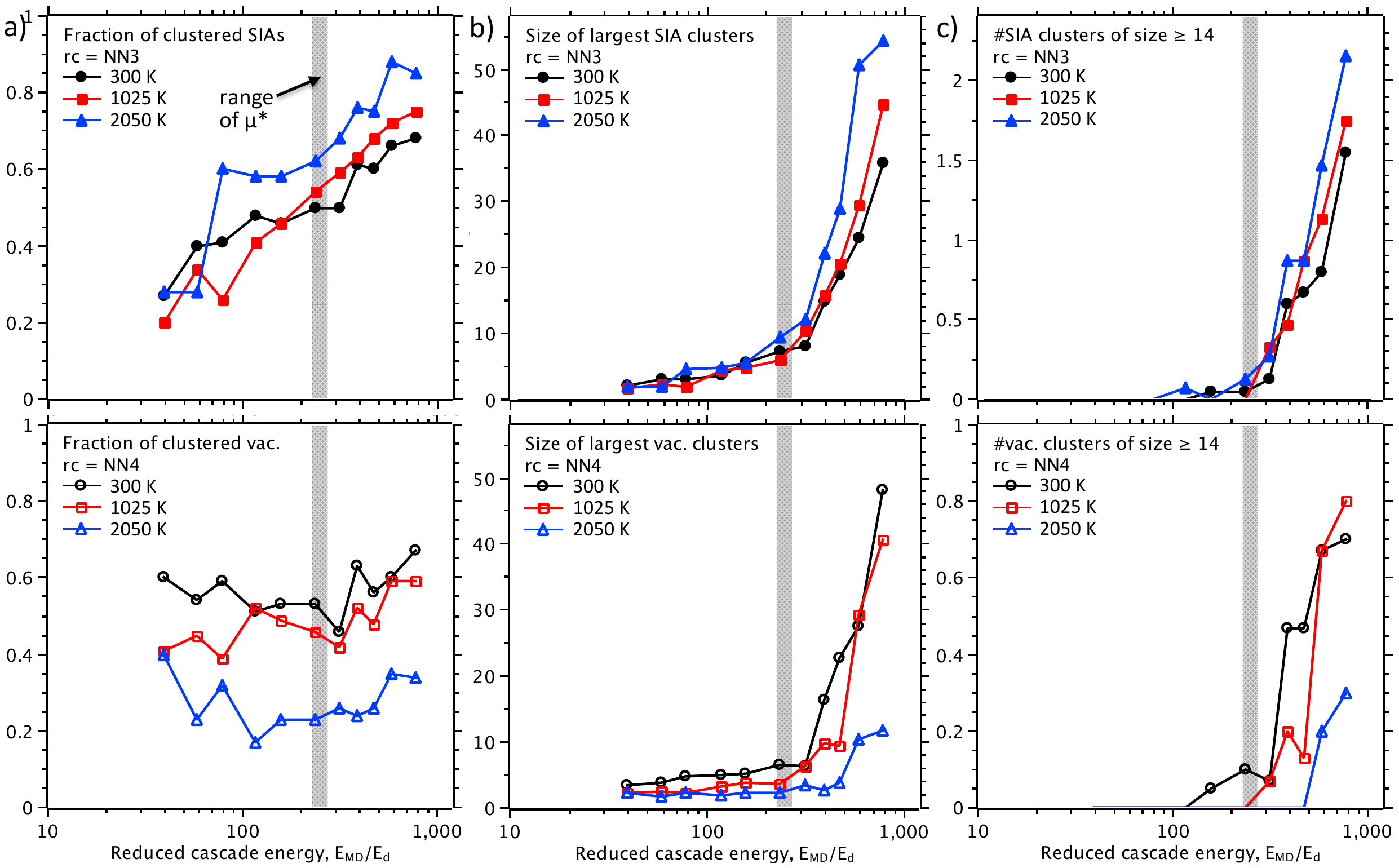}
\caption{ Energy dependence of the a) fraction of clustered defects, b) average size of the largest clusters and c) number of clusters with size $\ge$ 14 per cascade calculated with clustering cutoff up to NN3 for SIAs (top panels) and NN4 for vacancies (bottom panels).}
\label{figclustering}
\end{figure*}

During the MD simulations, occasional vacancy migrations are observed, but only at 2050 K. Such migrations are rare and do not alter the observed clustering behavior. Several graphs to illustrate the clustering trend as a function of energy and temperature are shown in Figure \ref{figclustering}. The plotted quantities are: the fraction of clustered defects, the average size of the largest clusters (\Smaxsia\ or \Smaxvac) and the number of large clusters per cascade (\Nlargesia or \Nlargevac). To calculate \Smaxsia\ or \Smaxvac, the size of the largest cluster from each simulation run is taken and then averaged over the number of runs. The minimum size of a cluster that can be resolved in a transmission electron microscope studies is a debatable matter. For the sake of our discussion, we consider a large cluster as one with size $\ge$ 14 (here, 14 signifies the sum of the first and second nearest neighbors in a tungsten lattice).

The clustered SIA fraction increases as a function of energy and exhibits a somewhat concave downward slope. At lower temperatures up to 1025 K, the fractions are still $<$ 0.5 for energies below the transition energy (shown by a vertical gray bar in Figure \ref{figclustering}). Above the transition, the fractions are between 0.5 and 0.7. Data at 2050 K shows a markedly larger fraction of clustered SIA with a value between 0.6 and 0.85. Evidently the higher fraction of clustered SIAs at higher temperatures is due to the migration and binding properties of the SIAs. While the clustered SIA fraction exhibits a concave curve, the \Smaxsia\ plots show a convex slope. Below the transition, \Smaxsia\ is limited up to five atoms. However, above the transition \Smaxsia\ increases rapidly and reaches 35, 45 and 55 atoms at 300 K, 1025 K and 2050 K, respectively. When we count the number of large clusters, there are practically no large clusters below the transition, however large clusters readily form at energies above the transition (Figure \ref{figclustering}c). This distinct increase in the probability of large SIA cluster formation is observed at all temperatures and indicates a correlation between the transition energy and the onset of large SIA cluster formation \cite{CascadeTransition}.

Temperature has an opposite effect on the fraction of vacancy clustering compared to SIA clustering. The fraction of clustered vacancies at 2050 K is about half of that at the lower temperatures of 300 K and 1025 K. Plots of \Smaxvac\ show that the average size of the largest vacancy clusters reaches about 40-50 vacancies at 300 K and 1025 K, but this value drops to around 10 at 2050 K. This indicates that while large vacancy clusters are still commonly found at temperature up to 1025 K, they are substantially reduced at the higher temperature of 2050 K.

\begin{figure*}[htbp]
\centering
\includegraphics[width=0.95\textwidth]{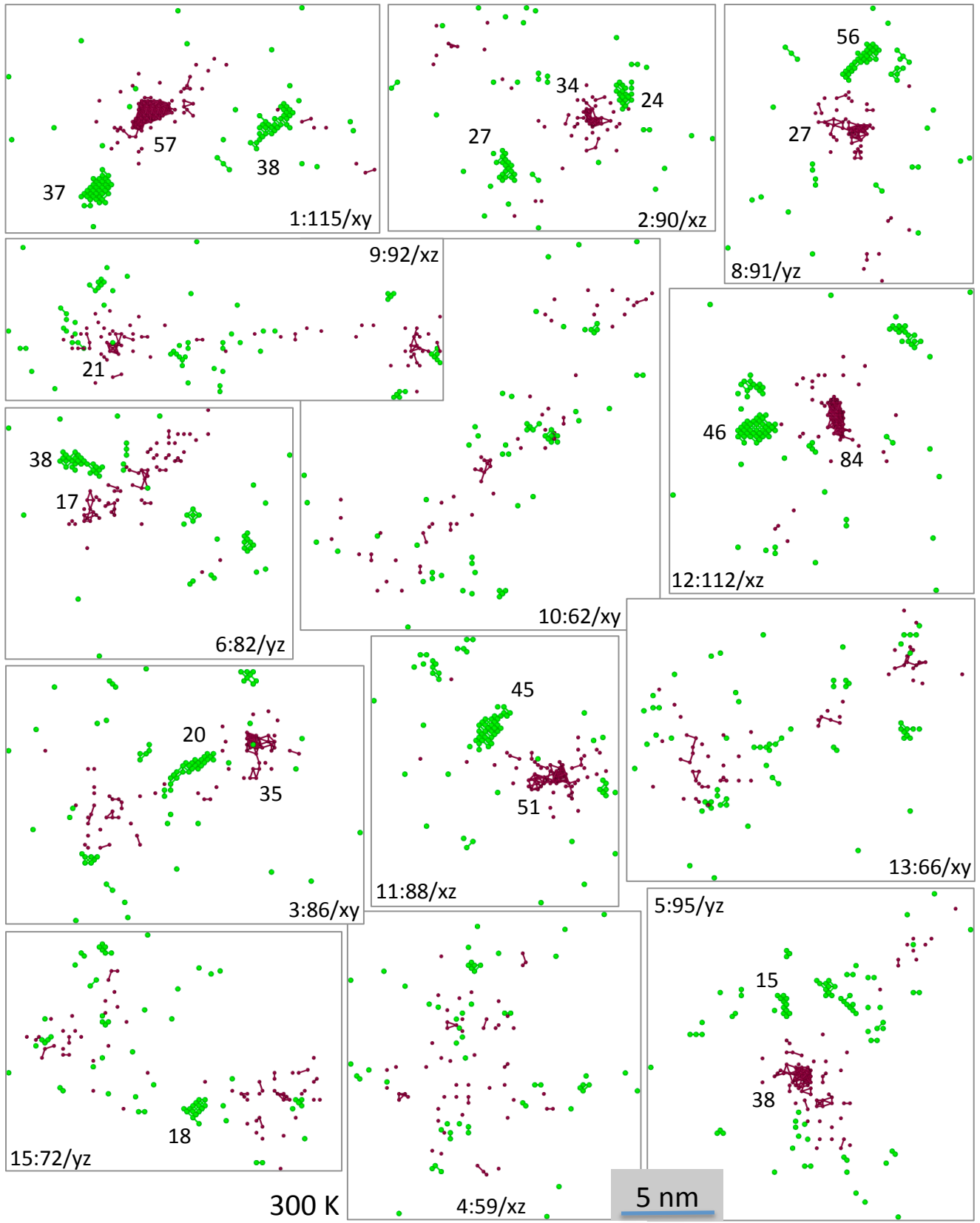}
\caption{(color online) Snapshots of surviving SIAs and vacancies plotted in gray and black dots (green and red in online version), respectively, for 75-keV cascade at 300 K, 13 out of 15 runs are shown. The label's format is run:\NF/view-projection. Clusters are determined with a cutoff up to NN3 and NN4 for SIAs and vacancies, respectively. The size of large clusters (size $\ge$14) is shown.}
\label{figsnapshot300K}
\end{figure*}

\begin{figure*}[htbp]
\centering
\includegraphics[width=0.95\textwidth]{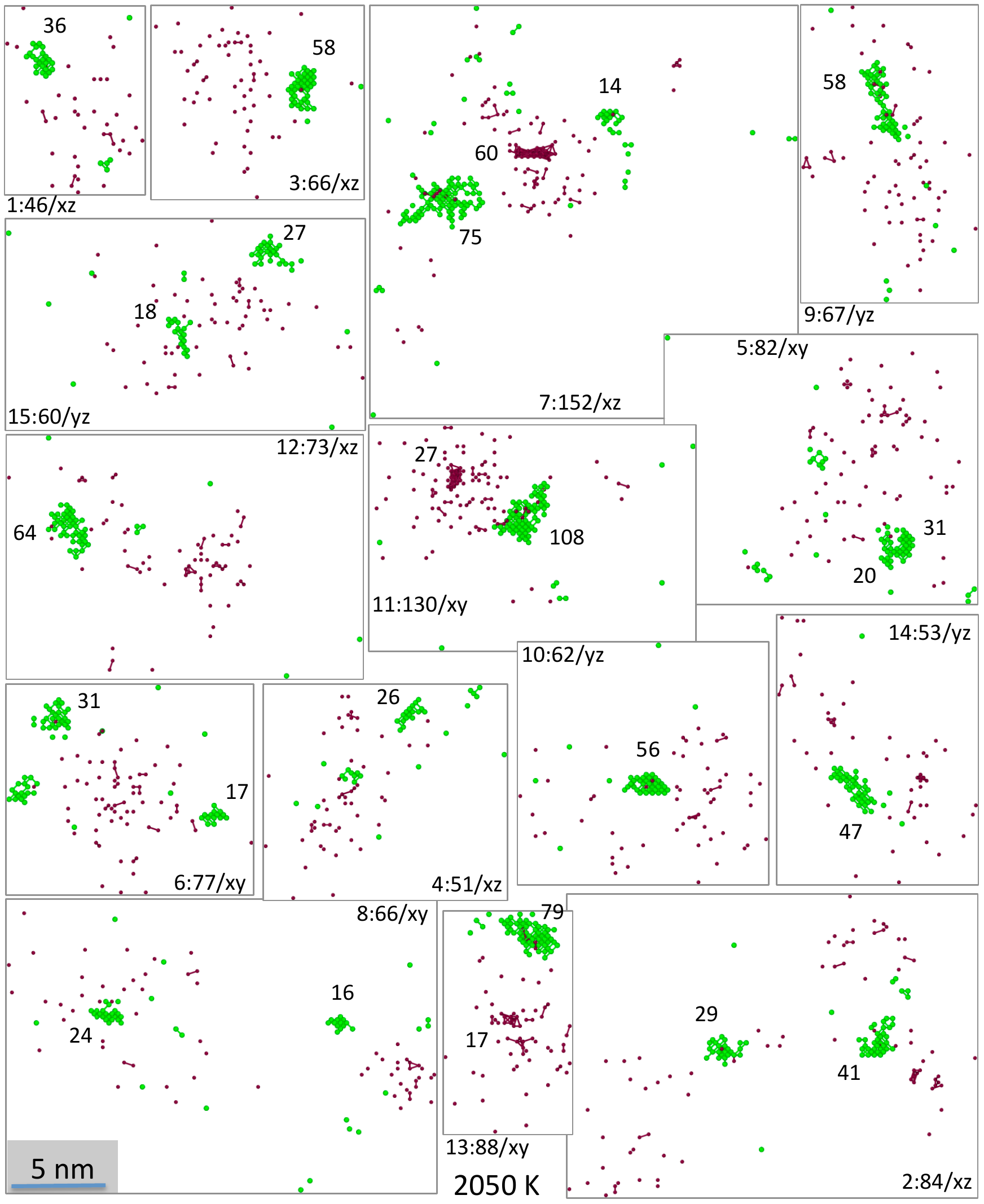}
\caption{(color online) Snapshots of surviving SIAs and vacancies plotted in gray and black dots (green and red in online version), respectively, for 75-keV cascade at 2050 K, all 15 runs are shown. The label's format is run:\NF/view-projection. Clusters are determined with a cutoff up to NN3 and NN4 for SIAs and vacancies, respectively. The size of large clusters (size $\ge$14) is shown.}
\label{figsnapshot2050K}
\end{figure*}

To better illustrate the effect of temperature on SIA and vacancy clustering, snapshots of surviving defects from 75-keV cascades at 300 K and 2050 K are depicted in Figures \ref{figsnapshot300K} and \ref{figsnapshot2050K}, respectively. Self-interstitial atoms or vacancies are plotted in gray or black dots (green or red in the online version). The label on each projected snapshot describes the particular simulation condition as follows. For example, the label 1:115/xy indicates that the snapshot is for simulation run-1 that produces \NF\ = 115 and the view is projected on the xy-plane. Large clusters are labeled with their size. Note that at this energy, there are 15 simulation runs. All snapshots at 2050 K are shown in Figure \ref{figsnapshot2050K}. For the 300 K data, only 13 out of 15 runs can be included in Figure \ref{figsnapshot300K} due to space limitation. At 300 K, the snapshots show that i) large vacancy clusters appear in simulations where large SIA clusters also form, ii) out of 15 simulations, both large vacancy clusters and SIA clusters form in 10 runs, and iii) single SIAs and mono vacancies appear in approximately similar proportion and spatially delocalized in a similar manner. On the other hand, at 2050 K the snapshots show that the SIAs belonging to large SIA clusters constitute the majority of their size distribution, in fact large clusters form in all simulations, while for vacancies, mono vacancies constitute the clear majority of the vacancy size distribution, with large clusters forming in only 3 runs. Snapshots at 1025 K look similar to those at 300 K with both large vacancy and SIA clusters forming in 10 simulations. The 1025-K snapshots are only included in the Supplemental Document \cite{supplemental} for brevity.

\subsection{Interstitial and vacancy loops}
In this study, all vacancy clusters of size $<$ 50 exhibit a three-dimensional (3D) shape (cavity). Larger clusters can be found as either cavities or platelets (loops). From all simulations, the number of vacancy loops at 300, 1025 and 2050 K is 13, 7 and 1, respectively. All vacancy loops appear to be $<$100$>$\{100\} loops. For SIAs, clusters of size $<$ 30 are all found as 3D SIA clusters. Larger clusters may form 3D clusters and loops. Considering only the SIA clusters of size $\ge$ 30, the partition of SIA clusters into loops and 3D clusters is given in Table \ref{tableSIAloops}. At 300 and 1025 K, the majority of these clusters are found as either \onehalf$<$111$>$\{111\} or \onehalf$<$111$>$\{110\} loops. On the other hand, the majority of the SIA clusters at 2050 K are 3D clusters. In MD cascade simulations of bcc metals, $<$100$>$ SIA loops are rarely observed. In this study, several $<$100$>$\{110\} SIA loops are observed at 1025 K and 2050 K. There are no $<$100$>$\{100\} SIA loops observed. In Table \ref{tableSIAloops}, several loops are identified as ÒmixedÓ loops. A mixed loop exhibits a \onehalf$<$111$>$ loop in one part and a $<$100$>$ loop in another part. Note that, each of these parts must constitute less than \twothird\ of the size for the loop to be determined as a mixed loop.

\begin{table*}[htbp]
\caption{Partitioning of SIA clusters of size $\ge$ 30 into loops and 3D clusters. 'Mixed loop' denotes a mixture of \onehalf$<$111$>$ and $<$100$>$ loops. '3D' denotes a 3D cluster of $<$111$>$ crowdions. 'Mixed 3D' denotes a mixture of $<$111$>$ and $<$100$>$ crowdions}
\begin{center}
\begin{tabular}{cccccccc}
\hline \hline
 & \EMD\ & \onehalf$<$111$>$\{111\} &  \onehalf$<$111$>$\{110\} & $<$100$>$\{110\} & Mixed loop & 3D & Mixed 3D\\
 & (keV)  &                                 &                           &            &     &       & \\
\hline
300 K & 100 & 9 & 4 & 0 & 0 & 1 & 0\\
            &  75  & 4 & 2 & 0 & 0 & 0 & 0\\
            &  60  & 2 & 1 & 0 & 0 & 1 & 0\\
            &  50  & 0 & 0 & 0 & 1 & 0 & 0\\
        &  Total  & 15 & 7 & 0 & 1 & 2 & 0\\
\hline
1025 K & 100 & 2 & 5 & 1 & 1 & 4 & 0\\
            &  75  & 2 & 2 & 1 & 1 & 0 & 0\\
            &  60  & 0 & 1 & 0 & 0 & 0 & 0\\
            &  50  & 0 & 0 & 1 & 0 & 2 & 0\\
        &  Total  & 4 & 8 & 3 & 2 & 6 & 0\\
\hline
2050 K & 100 & 0 & 0 & 1 & 1 & 17 & 2\\
            &  75  & 0 & 0 & 1 & 1 & 10 & 0\\
            &  60  & 0 & 1 & 0 & 0 & 3 & 1\\
            &  50  & 0 & 0 & 0 & 0 & 5 & 0\\
        &  Total  & 0 & 1 & 2 & 2 & 35 & 3\\
\hline \hline
\end{tabular}
\end{center}
\label{tableSIAloops}
\end{table*}

\begin{figure*}[htbp]
\centering
\includegraphics[width=0.80\textwidth]{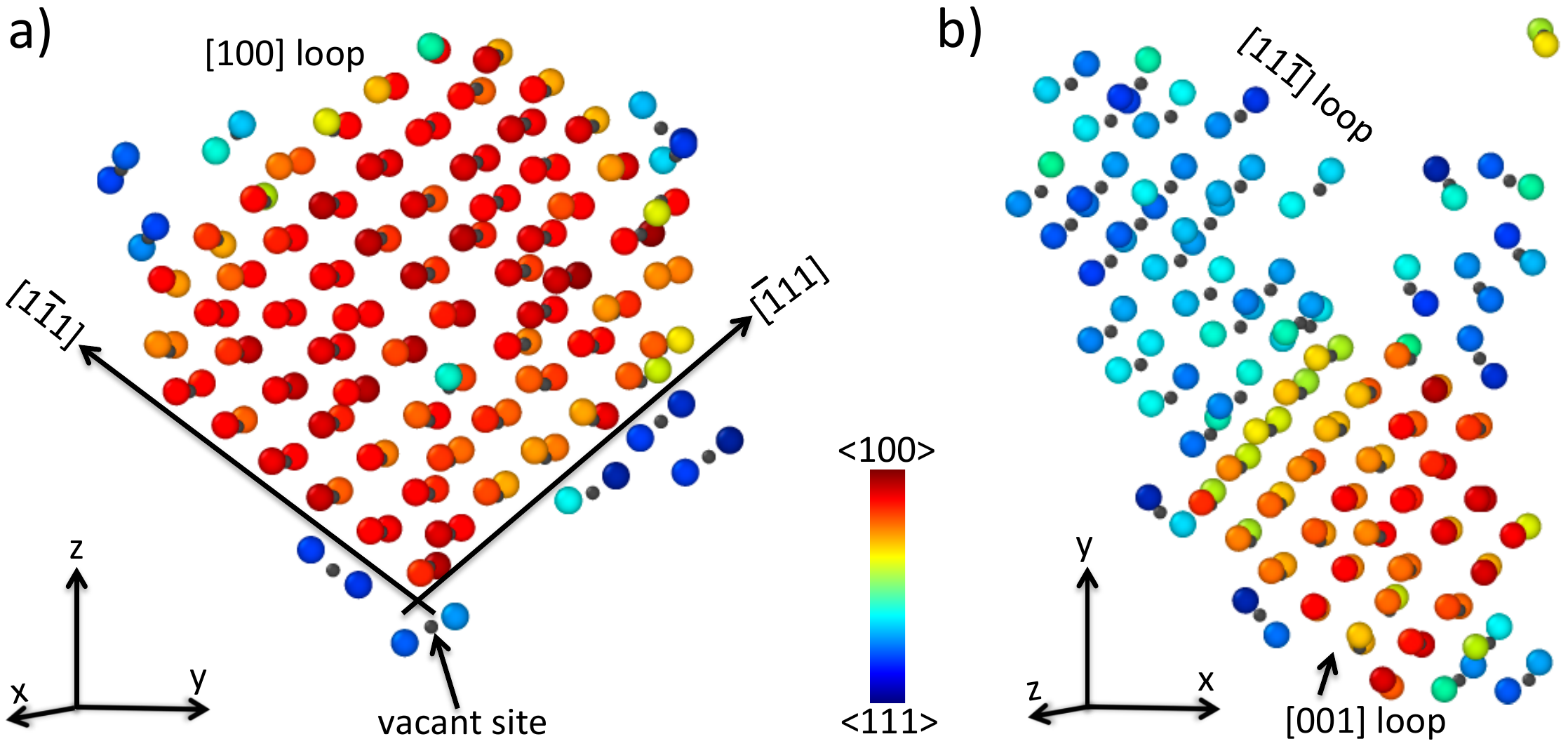}
\caption{(color online) Snapshots of a) [100](110) SIA loop of size 75 and b) mixed SIA loop of size 68, both obtained from 75-keV cascades at 1025 K.}
\label{figSIAloops}
\end{figure*}

Figure \ref{figSIAloops}a shows a [100](110) SIA loop of size 75 formed at 1025 K with a 75-keV PKA. The atoms in dumbbell configurations are colored based on their orientation. Red atoms are part of [100] dumbbells while blue atoms are part of $<$111$>$ dumbbells. During the formation of this loop, \onehalf$<$111$>$ SIA loops were not observed. It appears that the $<$100$>$ SIA loop is formed directly during the cooling of the cascade melt from a collection of interstitial atoms. All $<$100$>$ SIA loops observed in this study have a shape of a parallelogram (or a rhombus), not a rectangle. The edges of the parallelogram are oriented along $<$111$>$ on one of the \{110\} planes. Therefore, these edges form a set of 109.47$^\circ$ and (180$^\circ$-109.47$^\circ$) angles. Figure \ref{figSIAloops}b shows a mixed SIA loop, consisting of a [11\={1}] and a [001] loop. This loop is also formed at 1025 K but from a different 75-keV cascade compared to the loop shown in Figure \ref{figSIAloops}a. In Figure \ref{figSIAloops}b, both of the [11\={1}] and [001] loop parts are formed at an approximately the same time during the cascade cooling. It appears that both loops maintain their Burgers vector until the end of the MD run, resulting in a mixed SIA loop. Presumably, a much longer time is needed to see whether this loop will eventually transform into a $<$100$>$ or a \onehalf$<$111$>$ loop. In Fe, it was previously shown using the {\small SEAKMC} code that such transformation could occur on the microsecond time scale \cite{XuPRL13}.

\section{Discussion}

It is commonly understood that the effect of temperature on defect production is small but clear \cite{BaconJNM00}. Several factors that reduce the number of defects at higher temperature in MD simulations include higher defect mobilities, a longer life time of the cascade melt and the fact that the cascade itself tends to be more compact. As shown by the fit lines (see Figure \ref{figNF}a), the effect of temperature on the number of defects can be quantified more reliably. In Region 1, the number of defects decreases $\sim$20$\%$ from 300 K to 1025 K and $\sim$5$\%$ from 1025 K to 2050 K. One may consider that these variations are only a matter of annealing time scaling. However, our OKMC simulations that are described in the companion articles \cite{companion} suggest that the differences remain significant if the material contains microstructures that can capture SIAs migrating away from the cascade region.

The number of defects decreases only slightly from 1025 K to 2050 K. However, as described above, vacancy clustering is distinctly different between these two temperatures, i.e. large vacancy clusters are practically absent at 2050 K. Conversely, more SIA clustering occurs at the higher temperature (large SIA clusters form in every simulation run at 2050 K). The opposite effect of temperature on vacancy versus SIA clustering leads to a distinct asymmetry in the defect clustering behavior and the asymmetry increases with temperature. Such an asymmetry has an important implication on the long-term survival of those defects. The larger the asymmetry, the lower the probability for recombination within the aging of a single displacement cascade. Evidently, the ratios of kinetic parameters between vacancies and SIAs also influence the recombination outcome. Indeed, the OKMC simulations show that the fraction of surviving defects (i.e. \NF\ from OKMC simulations divided by that from MD results) is higher at 2050 K as compared to that at 1025 K \cite{companion}.

Through the combination of high PKA energies and temperatures, the profound effect of temperature on defect clustering asymmetry in tungsten is revealed. In the past, displacement cascade simulations in tungsten have been sparse and usually performed at 300 K and up to only 20 keV \cite{JuslinPhilMag10, FikkarJNM09, BjorkasNIMB09, ParkNIMB07}. Fikkar \etal. reported simulations for 10 keV at 523 K and 50 keV at 10 K \cite{FikkarJNM09}. On the other hand, a comprehensive set of cascade simulations was reported for Fe \cite{StollerBook12} that covered 100 K, 600 K and 900 K (which is 0.5 melting temperature of Fe). However, defect clustering in Fe is much less compared to that in W hence any temperature effect on clustering is not readily discernable. This is why the above asymmetry and its temperature dependence is not commonly known.

Recent experimental observations suggested that large interstitial clusters may be formed within individual cascades during {\it in situ} TEM with 150-keV self-ion irradiation in W \cite{YiPhilMag13}. The formation of such clusters is presumed to be an athermal process. Such clusters typically form within the first 0.2 ps of the cascade development when the destructive phase occurs \cite{CascadeTransition, CalderPhilMag10}. During this phase, the cascade energy is transferred from atom to atom much faster than the phonon-mediated dissipation allows, hence supersonic shock waves form. When regions of high atomic density associated with supersonic shock waves intersect, the excess atoms from the high-density region of one shock front are deposited into the low-density region (core) associated with the other shock waves and large interstitial clusters form. This process is some times referred to as in-cascade cluster formation to distinguish it from the clustering process due to thermal diffusion. We have verified that in-cascade SIA cluster formation applies at all temperatures in this study. Further, we note that subsequent capture reactions of SIA via thermal diffusion increase the size of these large clusters, as indicated by the trend of clustering data as a function of temperature.

When a large SIA cluster forms, it induces a depletion of atoms in other regions which potentially facilitates the formation of a large vacancy cluster. This scenario is reflected within the data at 300 K and 1025 K, i.e. large vacancy clusters are obtained in simulations where large SIA clusters are also observed. Nevertheless, the formation of a large vacancy cluster eventually depends on the interplay between the recrystallization of the atoms near the core and the motion of displaced atoms in the cascade melt \cite{NordlundPRB97}. In other words, the controlling factor seems to be the recrystallization rate. At a higher temperature, the cascade melt persists for a longer time and the displaced atoms are also more mobile. This combination increases the chance for those displaced atoms to refill the cascade core before the atoms surrounding the core recrystallize, hence a large vacancy cluster rarely forms at 2050 K.

\section{Conclusions}
In conclusion, we have presented displacement cascade simulations of tungsten over a wide range of PKA energies and temperatures. The data reveal a different regime of power-law energy-dependence of the defect production above a transition energy. At the lowest temperature (300 K), the transition occurs at 280$\times$\Ed, while at the highest temperature (2050 K), it occurs at 230$\times$\Ed. Below the transition, the number of surviving defects, \NF, scales with energy with an exponent of 0.74 at all temperatures, while above the transition, the exponent is 1.30 at 300 K which increases slightly to 1.36 at 2050 K. The more rapid increase of \NF\ above the transition is associated with the formation of large SIA clusters of size 14 or more.

For temperatures up to 1025 K, it is observed that the formation of large SIA clusters is accompanied by the formation of large vacancy clusters. The propensity of SIA clustering increases with temperature, however, the opposite is observed for vacancy clustering. This induces asymmetry in defect clustering, i.e. vacancy versus SIA clustering, which increases with temperature. A profound effect of temperature on such asymmetry is observed at 2050 K (i.e. 0.5 of melting temperature) at which practically no large vacancy clusters form while large SIA clusters form in all simulations. At 300 and 1025 K, vacancy clusters of size larger than 50 can be found as either $<$100$>$\{100\} loops or cavities. For SIA clusters larger than 30, the majority of them are found as either \onehalf$<$111$>$\{111\} or \onehalf$<$111$>$\{110\} loops at 300 and 1025 K. However, at 2050 K, 3D clusters constitute the majority of the SIA clusters. We note that rare $<$100$>$\{110\} SIA loops are observed at 1025 and 2050 K.

\section*{Acknowledgments}
This research is supported by the Office of Fusion Energy Sciences, US Department of Energy (\#DE-AC06-76RL0-1830). Computations were performed on Olympus cluster at Pacific Northwest National Laboratory (Fusion account). The authors would like to acknowledge the use of \ovito\ \cite{ovito} and \scidavis\ \cite{scidavis} softwares for  visualization and plotting.
\section*{References}


\begin{thebibliography}{10}
\bibitem{supplemental}
Suplemental document [please add a hypertext link to it].
\end{thebibliography}

\end{document}